\documentclass[twocolumn]{aastex631}
\usepackage{comment}

\usepackage{amsmath}
\usepackage{mathrsfs} 
\usepackage{graphicx}
\usepackage[normalem]{ulem}
\usepackage{soul}
\usepackage{xcolor}
\usepackage{macros_vdb}

\received{Aug 15, 2022}
\revised{March 5, 2023}
\accepted{March 22, 2023}

\shorttitle{Dynamical Heating of Dwarf Galaxies in a Fuzzy Dark Matter Halo}
\shortauthors{Dutta Chowdhury et al.}

\begin{document}

\title{On the Dynamical Heating of Dwarf Galaxies in a Fuzzy Dark Matter Halo}

\correspondingauthor{Dhruba Dutta Chowdhury}
\email{dhruba.duttachowdhury@mail.huji.ac.il}

\author[0000-0003-0250-3827]{Dhruba Dutta Chowdhury}
\affiliation{Department of Astronomy, Yale University, New Haven, CT-06511, USA}
\affiliation{Racah Institute of Physics, The Hebrew University of Jerusalem, Jerusalem 91904, Israel}

\author[0000-0003-3236-2068]{Frank C. van den Bosch}
\affiliation{Department of Astronomy, Yale University, New Haven, CT-06511, USA}

\author[0000-0002-8282-9888]{Pieter van Dokkum}
\affiliation{Department of Astronomy, Yale University, New Haven, CT-06511, USA}

\author[0000-0002-9497-9963]{Victor H. Robles}
\affiliation{Yale Center for Astronomy and Astrophysics, New Haven, CT-06520, USA}

\author[0000-0002-1249-279X]{Hsi-Yu Schive}
\affiliation{Department of Physics, National Taiwan University, Taipei 10617, Taiwan}
\affiliation{Institute of Astrophysics, National Taiwan University, Taipei 10617, Taiwan}
\affiliation{Center for Theoretical Physics, National Taiwan University, Taipei 10617, Taiwan}
\affiliation{Physics Division, National Center for Theoretical Sciences, Taipei 10617, Taiwan}

\author{Tzihong Chiueh}
\affiliation{Department of Physics, National Taiwan University, Taipei 10617, Taiwan}
\affiliation{Institute of Astrophysics, National Taiwan University, Taipei 10617, Taiwan}
\affiliation{Center for Theoretical Physics, National Taiwan University, Taipei 10617, Taiwan}

\begin{abstract}
  Fuzzy Dark Matter (FDM), consisting of ultralight bosons, is an intriguing alternative to Cold Dark Matter. Numerical simulations solving the Schr\"odinger-Poisson (SP) equation, which governs FDM dynamics, show that FDM halos consist of a central solitonic core (representing the ground state of the SP equation), surrounded by a large envelope of excited states. Wave interference gives rise to order unity density fluctuations throughout the envelope and causes the soliton to undergo density oscillations and execute a confined random walk in the central region of the halo. The resulting gravitational potential perturbations are an efficient source of dynamical heating. Using high-resolution numerical simulations of a $6.6 \times 10^{9} \Msun$ FDM halo with boson mass, $m_\rmb=8 \times 10^{-23} \eV$, we investigate the impact of this dynamical heating on the structure and kinematics of spheroidal dwarf galaxies of a fixed mass but different initial sizes and ellipticities. The galaxies are set up in equilibrium in the time-and-azimuthally averaged halo potential and evolved for $10 \Gyr$ in the live FDM halo. We find that they continuously increase their sizes and central velocity dispersions. In addition, their kinematic structures become strongly radially anisotropic, especially in the outskirts. Dynamical heating also causes initially ellipsoidal galaxies to become more spherical over time from the inside out and gives rise to distorted, non-concentric isodensity contours. These tell-tale characteristics of dynamical heating of dwarf galaxies in FDM halos can potentially be used to constrain the boson mass.
\end{abstract}

\keywords{Galaxy dynamics (591), Galaxy dark matter halos (1880), Gravitational interaction (669)}

\section{Introduction}
\label{sec:intro}

In recent years, the lack of evidence for supersymmetry at the Large Hadron Collider \citep[e.g.,][]{canepa19, adam22} and the null-detection of weakly interacting massive particles \citep[e.g,][]{liu17} have prompted a renewed interest in alternatives to the Cold Dark Matter (CDM) paradigm \citep[e.g.,][]{bertone05, feng10, tait12, arun17}. One of these alternatives that has received a lot of attention is Fuzzy Dark Matter (FDM), also known as Scalar Field Dark Matter or Wave Dark Matter, which postulates that dark matter consists of ultralight bosonic particles with masses in the range $10^{-23}\eV \lta m_\rmb \lta 10^{-19}\eV$ \citep[see][for recent reviews]{suarezetal14, hui17, niemeyer20, hui21}. The typical de-Broglie wavelength, $\lambda_{\rm db} = h /(m_\rmb\,\sigma)$, where $h$ is the Planck's constant, of such particles is astrophysically large (e.g., for $m_\rmb=10^{-22}\eV$, it is of order a kpc in halos with a velocity dispersion, $\sigma$, of the order of $100 \kms$). Since the occupation numbers of the bosonic FDM density field are huge, FDM behaves as a classical field, characterized by a wavefunction, $\psi$, that obeys the Schr\"odinger equation for a self-gravitating particle in a potential that relates to the density, $\rho = m_\rmb |\psi|^2$, via the Poisson equation. 

On large scales ($\lambda \gg \lambda_{\rm db}$), FDM behaves almost indistinguishably from CDM. On smaller scales, though, it differs from CDM in three distinct ways. First, on scales $\lambda \lta \lambda_{\rm db}$, the uncertainty principle gives rise to a large quantum pressure, which leaves its imprint on the matter power spectrum and on the halo mass function at the low mass end \citep[e.g.,][]{du17, kulkarni20, may21, may22, nori22, sipp22}. One may thus hope to constrain FDM with similar probes as are used to discriminate between CDM and Warm Dark Matter (WDM) models, such as the Ly-$\alpha$ forest \citep[e.g.,][]{armengaud17, kobayashi17, irsic17, nori19, rogers21} or the abundance of low mass satellite galaxies in the Milky Way \citep[e.g.,][]{nadler19, schutz20, benito20, banik21, nadler21}.

Second, FDM halos are predicted to have a structure that is different from the NFW density profiles \citep[][]{navarro97} of CDM halos. In particular, numerical simulations that solve the Schr\"odinger-Poisson (SP) equation show that FDM halos consist of a central soliton (representing the ground state of the SP equation), surrounded by an NFW-like envelope made up of excited states \citep[e.g.,][]{schive14a, schwabe16, mocz17, veltmaat18, schwabe21, huang22}. Unlike the NFW cusp, the soliton has a constant density core (dominated by quantum pressure), and early simulation results hinted at a tight scaling relation between halo mass and the central density of the soliton for a given boson mass \citep[e.g.,][]{schive14b, schwabe16}. Hence, stellar or gas-kinematics based inferences on the central density profiles of dark matter halos may be used to constrain FDM in general and the boson mass in particular \citep[see e.g.,][for some attempts along these lines]{marsh15, calabrese16, gonzalez-moralez17, bar18, wasserman19, burkert20, pozo20, safarzadeh20, hayashi21, pozo21, bar22, pozo23}. However, more recent simulations seem to suggest an appreciable amount of scatter in the properties of the soliton (e.g., central density, spatial extent) for a given halo and boson mass \citep[e.g.,][]{may21, nori21, chan22, zagorac22b}, which hampers our ability to constrain FDM this way. Another limitation is the hitherto unexplored effect of baryonic processes such as supernova feedback, which can alter the structural properties of the soliton and the surrounding envelope.

The third property of FDM that distinguishes it from CDM is that wave interference causes pronounced density and potential fluctuations in the NFW-like halo envelopes \citep[e.g.,][]{schive14a, hui17, duttachowdhury21, yavetz22, zagorac22a, liu22}. These fluctuations have an amplitude of order unity, a characteristic size of the order of the de-Broglie wavelength, and can be envisioned as `quasiparticles' of mass $M \propto \rho \lambda^3_{\rm db}$, with $\rho$ the local FDM density, that pop in and out of existence. The resulting granular nature of FDM halos can be probed with gravitational lensing, which in turn can be used to constrain the boson mass \citep[][]{laroche22, powell23}. In addition, the quasiparticles act as an efficient source of dynamical heating \citep[e.g.,][]{bar-or19, elzant20a, elzant20b}, which can cause, among others, a diffusion of stellar streams \citep[][]{amorisco18, dalal20} or a thickening of stellar disks \citep[][]{church19, chiang23}. The wave interference also impacts the soliton, which undergoes order-unity temporal oscillations in its density as well as a constrained random walk within the central region of the halo by of order its own extent \citep[][]{veltmaat18, schive20, duttachowdhury21, li21, zagorac22a,chiueh22}. The rapid time variability in the gravitational potential resulting from the wobbling and oscillating soliton is an additional source of dynamical heating. 

Dynamical heating due to the central soliton (and the quasiparticles) is likely to leave an observable imprint on galaxies or nuclear objects (e.g., central supermassive black holes and nuclear star clusters) that reside in FDM halos, which can ultimately be used to test and constrain FDM. For example, in \citet[][]{duttachowdhury21}, we investigated the impact of FDM potential fluctuations on the motion of nuclear objects. Using high-resolution numerical simulations of an FDM halo of virial mass $\sim 6.6 \times 10^{9} M_{\odot}$ and $m_\rmb=8 \times 10^{-23} \eV$, we demonstrated that nuclear objects, initially at rest at the soliton center, diffuse outwards with time. This outward diffusion continues until counteracted by dynamical friction \citep[see e.g.,][]{lancaster20, wang21, vicente22, vitsos22}, such that less massive objects diffuse out to larger radii. Hence, FDM models predict that (compact) nuclear star clusters and/or supermassive black holes should (on average) be offset from the centers of mass of their host galaxies. Interestingly, such offsets are not uncommon in dwarf galaxies \citep[e.g.,][]{binggeli00, cote06, shen19, reines20}. However, to what extent these data constrain the boson mass remains to be determined.

In addition to their outward diffusion, nuclear star clusters in FDM halos can also be disrupted if they are sufficiently diffuse. For example, \citet{marsh19} argue that the presence of a nuclear star cluster in Eridanus II implies that $m_\rmb \gtrsim 0.6-1 \times 10^{-19}\eV$, as for smaller boson masses, the temporal oscillations of the soliton would completely disrupt the fairly diffuse star cluster. However, \citet{chiang21} show that irrespective of $m_{\rmb}$, the time period of soliton oscillations is much larger than typical orbital timescales within the star cluster, such that it cannot be significantly heated via this mechanism. While the soliton oscillations are adiabatic, \citet{schive20} demonstrate that for a boson mass of $8 \times 10^{-23} \eV$, the soliton random walk would completely disrupt the star cluster in $\sim 1 \Gyr$ if Eridanus II were to be located in the field. But, given the galaxy's orbit around the Milky Way, the same study also shows that its outer NFW-like envelope is stripped off with time, greatly reducing the magnitude of the soliton random walk, which is the dominant heating mechanism. Consequently, the Eridanus II star cluster can survive for several Gyr.

Using the impact of the dynamical heating on galaxies residing in FDM halos, \citet[][]{dalal22} argue that the existence of the ultra-faint dwarf galaxies (UFDs) Segue 1 and 2 implies that $m_\rmb > 3 \times 10^{-19} \eV$. For smaller boson masses, they show that the dynamical heating would rapidly increase the sizes and velocity dispersions of the two UFDs to values much larger than observed. However, \cite{dalal22} use an approximate method to estimate the dynamical heating from the quasiparticles and do not consider the heating from the soliton. Although they argue that adding the soliton-specific heating mechanisms would only increase the actual heating rate, thereby further tightening their constraint, we point out the following potential caveats to their analysis. As Segue 1 and 2 are satellite galaxies of the Milky Way, their outer halo envelopes are likely to have been stripped off, causing the dynamical heating due to both the soliton and the quasiparticles to be significantly suppressed. This is true even if the outer envelopes are just partially stripped \citep[][]{schive20}. In addition, since the ratio of soliton mass to halo mass increases with increasing redshift and decreasing halo mass \citep[][]{schive14b}, if Segue 1 and 2 formed sufficiently early, perhaps even before reionization, their host halos may never have had much of a fluctuating envelope to begin with before they were accreted by the Milky Way's progenitor halo. Although somewhat contrived, these caveats emphasize the need to properly take into account mass assembly and tidal mass loss (in the case of subhalos such as the MW satellites) of FDM halos, either using cosmological simulations \citep[e.g.,][]{schive14a, nori22} or (approximate) analytical halo evolution models \citep[e.g.,][]{du17, du18, du23}, to derive a more robust constraint on the boson mass using observable galaxy properties.

In this paper, we examine the impact of FDM potential fluctuations and the resulting heating effect on isolated dwarf galaxies, i.e., galaxies that reside in their own dark matter halos rather than being satellites of a more massive system, using SP simulations. By employing the {\tt GAMER-2} code, we simulate the dynamics of dwarf galaxies of a fixed mass but different initial ellipticities and sizes in the same FDM halo as that used in \citet[][]{duttachowdhury21}. Each dwarf galaxy is initialized to be in equilibrium in the time-and-azimuthally averaged halo potential and subsequently evolved for $10 \Gyr$ in the fluctuating potential of our live FDM halo. The main goal of this paper is not to provide any quantitative constraints on the boson mass but rather to highlight the observable consequences of the dynamical heating of (spheroidal) dwarf galaxies in FDM halos. 

The paper is organized as follows. Section~\ref{sec:setup} describes our methodology and the different simulation setups, the results of which are presented in Section~\ref{sec:results}. We summarize and conclude in Section~\ref{sec:concl}.

\section{Simulation SetUp} 
\label{sec:setup}

\begin{figure*}
    \centering
    \includegraphics[width=0.95\textwidth]{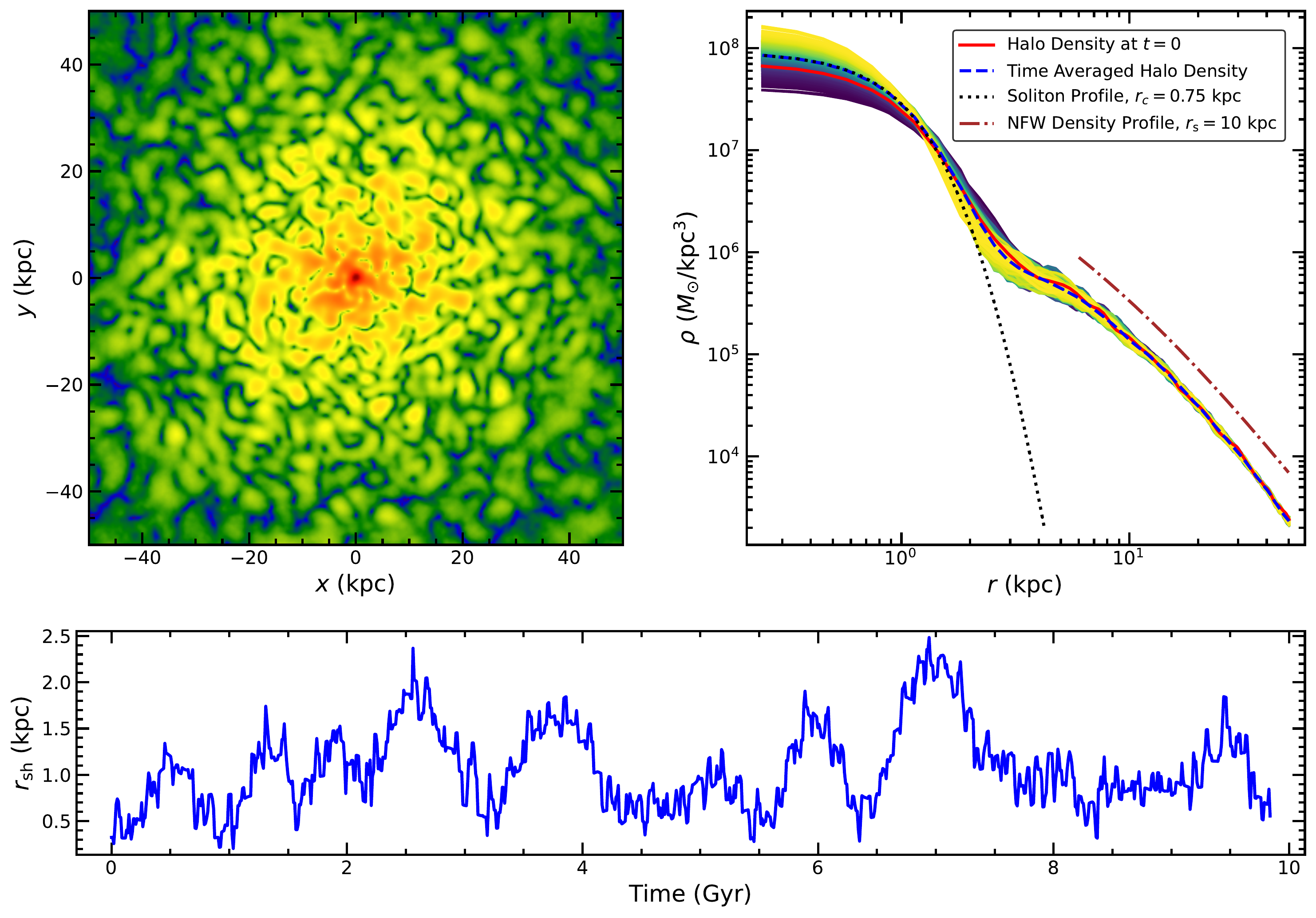}
    \caption{The top-left panel shows the density of our FDM halo at $t=0$ in a thin $(x,y)$-slice of dimensions $50 \kpc \times 50 \kpc$ centered on the maximally dense cell. The halo consists of a ground state, also known as the soliton (red, central nugget), surrounded by a much larger envelope of excited states that extensively interfere with one another, giving rise to the density fluctuations seen throughout the envelope. The red curve in the top-right panel shows the azimuthally averaged density profile at $t=0$, where $r=0$ corresponds to the location of the maximally dense cell. The other solid curves indicate the same but at different times, covering a period of $10 \Gyr$, at spacings of $0.01 \Gyr$, highlighting the order unity temporal oscillations in the soliton density. The curves are color-coded based on the central density, which shows that an increase in peak density is associated with a decrease in soliton size and vice-versa. The solid curves are averaged to give the dashed, blue curve, which depicts the time-and-azimuthally averaged halo density profile. It is soliton-like (core radius, $r_{\rmc}=0.75 \kpc$, dotted black curve) at small $r$, up to about $r_{\rm sol}=2.7 r_{\rmc}$, and NFW-like at large $r$ (scale radius, $r_{\rm s}=10 \kpc$, dot-dashed brown curve). Finally, the bottom panel depicts the soliton random walk, showing the spatial offset of the soliton center, defined as the location of the maximally dense cell, from the halo center of mass as a function of time.}
    \label{fig:halo_properties}
\end{figure*}

\begin{figure*}
    \centering
    \includegraphics[width=0.95\textwidth]{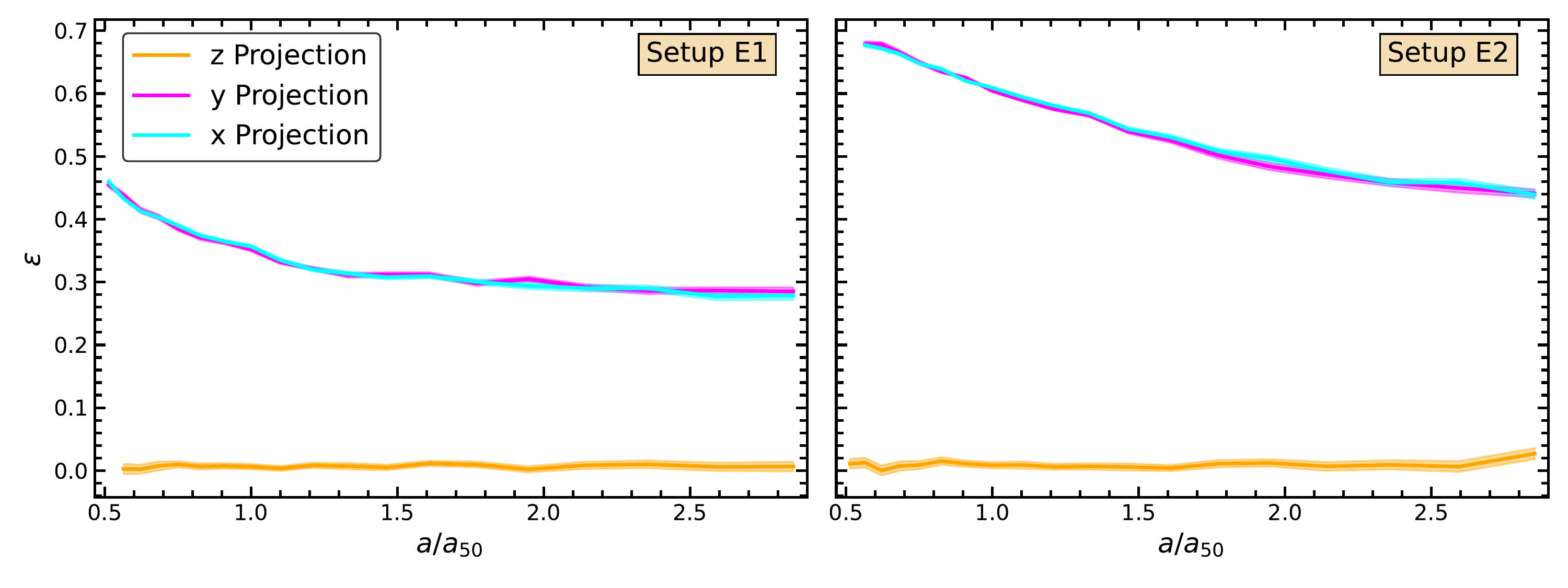}
    \caption{Left- and right-hand panels show the initial ellipticity profiles of the oblate galaxies in Setups E1 and E2, respectively, for the $x$ (cyan), $y$ (magenta), and $z$ (orange) projections. These are obtained by fitting elliptical isodensity contours to the projected particle distributions. For each setup, the ellipticity, $\varepsilon$, is plotted as a function of the semi-major axis, $a$, normalized by $a_{50}$, the semi-major axis of the isodensity ellipse that encloses $50 \%$ of the galaxy mass in projection. The solid curves and the associated envelopes (barely visible), respectively, indicate the best fits and the $1 \sigma$ error bounds.}
    \label{fig:initial_ellipticity}
\end{figure*}

Throughout this paper, we study the evolution of dwarf galaxies in one particular FDM halo, which has a virial mass of $\Mvir = 6.6 \times 10^{9} \Msun$. Its initial wavefunction is extracted from the redshift zero snapshot of a large, cosmological simulation of structure formation in a universe with a bosonic dark matter particle mass of $m_\rmb = 8 \times 10^{-23} \eV$ \citep[see][for details]{schive14a}. In order for it to attain equilibrium, the extracted halo is simulated in isolation for several Gyr at a uniform spatial resolution of $244 \pc$ within a box of dimensions $125 \kpc \times 125 \kpc \times 125 \kpc$ using the code {\tt GAMER-2} \citep{schive18}, which evolves the system by solving the SP equation. We adopt periodic boundary conditions when updating the wave function (i.e., the flow traveling across the right edge will re-enter the simulation domain from the left edge) and isolated boundary conditions (i.e., the potential is zero at infinity) when computing the gravitational potential. A fixed time step of $6.3 \times 10^5 \yr$ is used, which is motivated by stability considerations of the kinetic and potential energy operators \citep[see][for details]{schive14a}. The virial radius of our halo is $r_{\rm vir} \sim 50 \kpc$, indicating that the simulation box encloses the entire virial volume but not much more. In what follows, we take this evolved, equilibrium state of the halo to correspond to $t=0$.

\subsection{Halo Structure and Evolution}
\label{halo_prop}

The top-left panel of Figure~\ref{fig:halo_properties} shows a thin $(x,y)$-slice of the density of the halo at $t=0$ centered on the maximally dense cell. The halo consists of a ground state, also known as the soliton, indicated by the red, central nugget. The soliton is surrounded by a much larger envelope of excited states that extensively interfere with one another, giving rise to the density fluctuations seen throughout the envelope. The characteristic features of these density fluctuations have been studied in detail in \citet[][]{duttachowdhury21}. They have order unity variance, independent of location, and a typical diameter of order the de-Broglie wavelength \citep[see also][]{bar-or19, elzant20a, chavanis20}.

The red curve in the right-hand panel of Figure~\ref{fig:halo_properties} depicts the azimuthally averaged density profile for this snapshot, where $r=0$ is the location of the maximally dense cell. In order to illustrate some interesting features of this halo, we evolve it for an additional $10 \Gyr$. The solid curves, varying in color from yellow to violet with decreasing central density, show the azimuthally averaged density profiles at intervals of $0.01 \Gyr$ over this time period. Note that the soliton undergoes order unity temporal oscillations in density \citep[see also][]{veltmaat18, duttachowdhury21, chiang21}. In addition, when the soliton becomes denser, it shrinks in size and vice-versa. This behaviour is in tune with the $\rho_{0}-r_{\rmc}$ scaling relation for solitons \citep[see e.g.,][]{seidel90, guzman06}, given by
\begin{equation}
\rho_0 = 1.95 \times 10^7 \Msun \kpc^{-3} \, m_{22}^{-2} \, \left( \frac{r_\rmc}{\kpc} \right)^{-4} \,,
\label{core_radius}
\end{equation} 
where $m_{22} = m_\rmb/(10^{-22}\eV)$, $\rho_{0}$ is the peak soliton density, and $r_{\rmc}$ is the core radius, defined as the radius where the soliton density drops to half of its maximum value. 

The dashed, blue curve in Figure~\ref{fig:halo_properties} indicates the time-and-azimuthally averaged density profile of our FDM halo, which is obtained by averaging all
the solid curves. For comparison, the dotted, black curve highlights the universal soliton profile of \cite{schive14a}, given by
\begin{equation}
    \rho_{\rm sol}(r) = \frac{\rho_0}{[1+0.091\ (r/r_\rmc)^{2}]^8} \ ,
    \label{sol_prof}
\end{equation}
with $r_{\rmc}=0.75 \kpc$ and $\rho_{0}$ determined using Equation~\ref{core_radius}. The time-and-azimuthally averaged density profile is consistent with the dotted, black curve at small radii but ceases to be soliton-like at around $2.7 r_{\rmc}=2 \kpc$, which we define as the soliton radius, $r_{\rm sol}$. Following a transition region, it becomes NFW-like at large radii ($r \gtrsim 5 \kpc$). A reference NFW profile \citep{navarro97} with scale radius, $r_{\rms}=10 \kpc$ is shown with the dot-dashed, brown curve, which is shifted upwards from the halo profile for clarity. 

In addition to exhibiting temporal oscillations in density, the soliton also moves around with respect to the halo center of mass, akin to a random walk \citep[see also][]{schive20, duttachowdhury21}. The spatial offset of the soliton center (defined as the location of the maximally dense cell) from the halo center of mass, $r_{\rm sh}$, as a function of time is depicted in the bottom panel of Figure~\ref{fig:halo_properties}. Note that the radial extent of the soliton's random walk is comparable to its own size, i.e., $r_{\rm sh} \lta r_{\rm sol}$. As shown in \citet{li21}, both the temporal oscillations and the random walk of the soliton arise from its interference with the excited states that make up the surrounding envelope. 

\begin{table*}
   \centering
    \begin{tabular}{| c | c | c |}
    \hline
    Simulation Setup & scale factor ($p$) & initial size ($a_{\rm gal}$) \\
    \hline
    S1 & $1.0$ & $0.3\kpc$ \\
    S2 & $1.0$ & $0.7\kpc$ \\
    S3 & $1.0$ & $1.5\kpc$ \\
    E1 & $0.5$ & $0.7\kpc$ \\
    E2 & $0.3$ & $0.7\kpc$ \\
    \hline
    \end{tabular}
    \caption{Overview of the various dwarf galaxy simulation setups discussed in the text. The scale factor, $p$, listed in Column 2, is the factor by which the positions and velocities of star particles along the $z$-direction are scaled to produce an oblate, flattened system. The galaxies in Setups S1, S2, and S3 are spherically symmetric Plummer spheres, and Column 3 lists their scale radii. For Setups E1 and E2, in which the galaxies are initialized as oblate spheroids, Column 3 lists the scale radius of the Plummer sphere from which they were generated, as described in the text. All setups comprise of 10 random realizations, as described in the text, and are evolved in the same FDM halo. Each galaxy is always represented with $10^{6}$ particles and has a total mass of $M_{*}=10^{6} M_{\odot}$.}
    \label{tab:sims}
\end{table*}

\subsection{Dwarf Galaxy Initial Conditions}
\label{galaxy_ics}

The random walk of the soliton, its oscillations, and the density fluctuations in the outer envelope are all sources of gravitational potential fluctuations. To investigate the impact of these potential fluctuations on dwarf galaxies, we simulate the evolution of both spherical and elliptical dwarf galaxies within our FDM halo.

Our spherical dwarfs are initially set up as Plummer spheres of mass $M_{\rm gal}=10^{6} M_{\odot}$ and three different scale radii, $a_{\rm gal}=0.3 \kpc$ (Setup S1), $0.7 \kpc$ (Setup S2), and $1.5 \kpc$ (Setup S3), always in equilibrium in the time-and-azimuthally averaged halo potential, $\Phi_\rmh(r)$, which is obtained from the time-and-azimuthally averaged halo density profile (dashed, blue curve in the right-hand panel of Figure~\ref{fig:halo_properties}) using the Poisson equation. Given $\Phi_\rmh(r)$ and $\rho_{\rm gal}(r)$, the Plummer density profile of the dwarf galaxy in a particular setup, the Eddington inversion formula \citep[][]{binney08} is used to obtain the corresponding ergodic distribution function, $f(E)$, where $E$ is the energy of a star particle. When computing $f(E)$, we account for the self-gravity of the dwarf galaxy but note that this is negligible compared to the halo potential. The galaxies in each setup are represented with $10^6$ particles of mass $m=1 \Msun$, whose positions and velocities are drawn from the corresponding $f(E)$. 

To create elliptical dwarfs, we take the spherical galaxy in Setup S2 and scale the positions and velocities of all star particles along the $z$ axis by a factor $p$. We choose two different values of $p$, $0.5$ (Setup E1) and $0.3$ (Setup E2), and using {\tt GAMER-2}, evolve the resulting oblate spheroids in the time-and-azimuthally averaged halo potential for several Gyr, allowing them to attain equilibrium. The equilibrated galaxies serve as initial conditions for our elliptical dwarfs. Their ellipticity profiles are shown in the left- (Setup E1) and right-hand (Setup E2) panels of Figure~\ref{fig:initial_ellipticity}. These are obtained by fitting elliptical isodensity contours to the projected particle distributions using the publicly available {\it photutils} package \citep{bradley20}. The resulting ellipticities, $\varepsilon$, are plotted as a function of the semi-major axes, $a$, normalized by $a_{50}$, which is the semi-major axis of the isodensity contour that encloses $50 \%$ of the total stellar mass in projection.

Table~\ref{tab:sims} summarizes the five different dwarf galaxy setups used in this paper. For each setup, 10 different realizations are created by introducing the galaxy within the FDM halo at different instants of time, $t_{0}$, such that the center of mass of the galaxy is always coincident with the soliton center and the velocity of its center of mass is equal to the soliton velocity. All realizations are evolved using {\tt GAMER-2} for $10 \Gyr$ with the same box-size and boundary conditions as used in the dark matter-only run. While the base spatial resolution is still $244 \pc$, we now add one level of refinement, such that grid patches (each with $8^3$ cells) are adaptively refined if they have more than 1000 star particles, yielding a maximum spatial resolution of $122 \pc$. All particles and grid patches on the same refinement level share the same time step, which is adaptively determined according to the criteria laid out in \citet[][]{schive18} and is different for different levels. We have verified that our results are converged by also running several simulations with an additional level of refinement, down to a maximum spatial resolution of $61 \pc$, the results of which are indistinguishable from those of our fiducial runs presented below.

\section{Results} 
\label{sec:results}

Before describing our main findings, we present an example showing the evolution of a representative dwarf galaxy taken from one of our 50 simulations ($5$ setups $\times 10$ realizations each). We then discuss and compare the statistical evolution of some key properties of the dwarf galaxies in the different simulation setups.

\subsection{A Specific Example}
\label{sec:example}

\begin{figure*}
    \centering
    \includegraphics[width=0.95\textwidth]{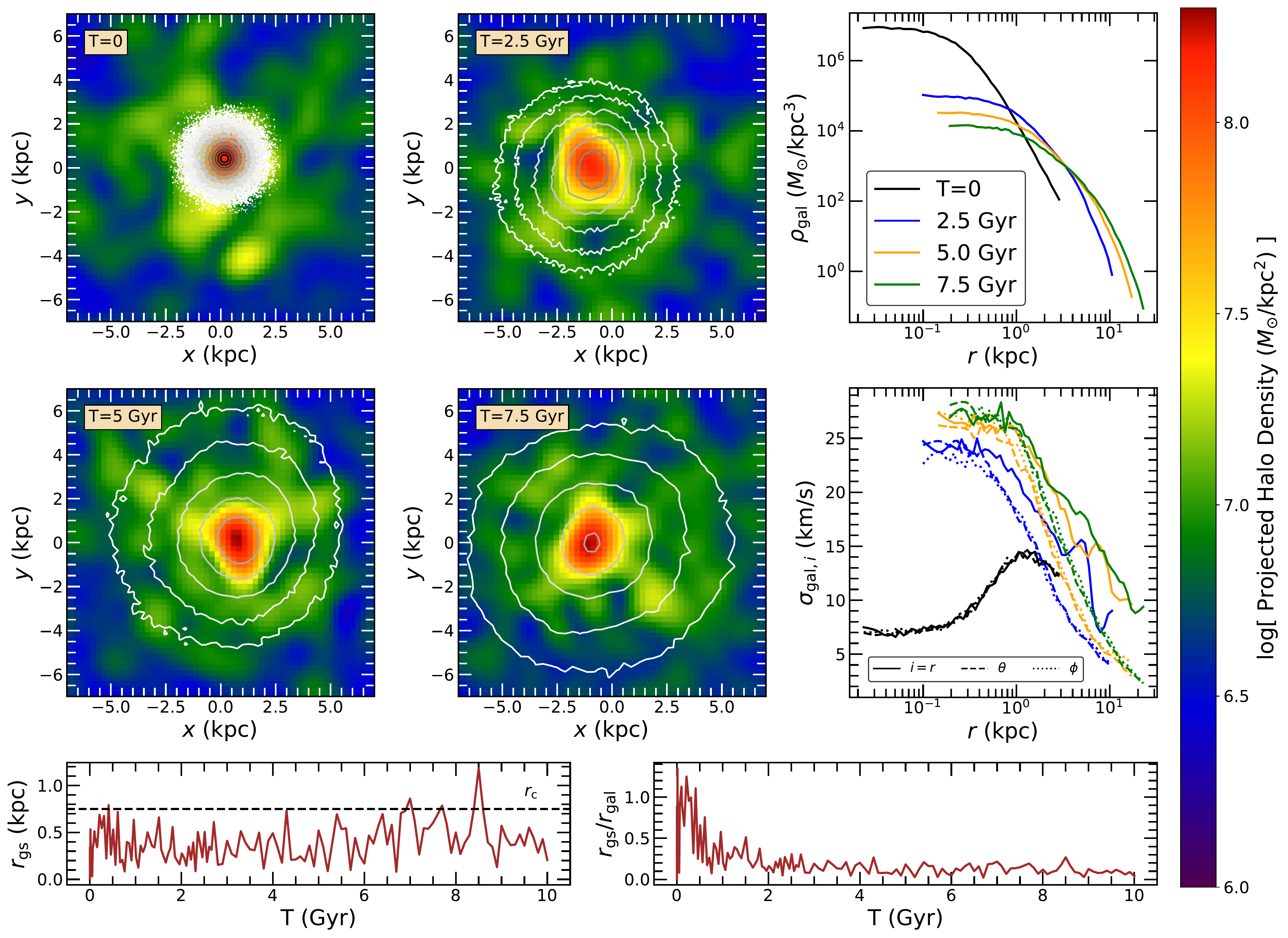}
    \caption{The temporal evolution of a dwarf galaxy in one randomly selected realization of Setup S1. The four images in the upper-left corner depict the projected densities of the halo and the galaxy in the $x$-$y$ plane at four different times, $T$, as indicated. In each image, centered on the center of mass of the halo, the color map shows the projected density of the halo and is overplotted with isodensity contours (grayscale) of the projected distribution of star particles. The contours cover three orders of magnitude in surface density, ranging from $2 \times 10^{6} \Msun \kpc^{-2}$ (black) to $2 \times 10^3 \Msun \kpc^{-2}$ (white). At $T=0$ (upper-left panel), these contours overlap due to the compactness of the initial galaxy. For the same values of $T$, the top and central right-hand panels, respectively, show the azimuthally-averaged 3D density, $\rho_{\rm gal}$, and 1D velocity dispersion, $\sigma_{{\rm gal},i}\;(i=r, \theta, \phi)$, of the galaxy, both as a function of the distance, $r$, from its center of mass. Finally, the bottom left- and right-hand panels depict the temporal evolution in the offset, $r_{\rm gs}$, of the galaxy center of mass from the soliton center, and the ratio of $r_{\rm gs}$ to the galaxy's 3D half-mass radius, $r_{\rm gal}$, respectively. The horizontal, dashed line indicates the soliton core radius, $r_\rmc$, and is shown for comparison. See text for details.}
    \label{fig:example}
\end{figure*}

Figure~\ref{fig:example} highlights the evolution of a dwarf galaxy in one randomly selected realization of Setup S1. The four images in the left-hand and middle panels of the top two rows show the projected density of the halo and the galaxy in the $x$-$y$ plane at four different times, $T=t-t_0$, as indicated. Here $t_0$ is the time when the galaxy is introduced into the FDM halo. In each image, which is centered on the center of mass of the halo, the projected density of the halo is depicted with a color map, while the projected density of the galaxy, $\Sigma_{\rm gal}$, is indicated by its isodensity contours in grayscale. For the same values of $T$, the top and central right-hand panels plot, respectively, the azimuthally-averaged three-dimensional (3D) density, $\rho_{\rm gal}$, of the galaxy and its one-dimensional (1D) velocity dispersion, $\sigma_{{\rm gal},i}$, as a function of the distance, $r$, from its center of mass. Here, $i=r$, $\theta$, $\phi$ are the usual spherical coordinates. Finally, the bottom left- and right-hand panels depict the temporal evolution in the offset, $r_{\rm gs}$, of the galaxy center of mass from the soliton center, and the ratio of $r_{\rm gs}$ to $r_{\rm gal}$, respectively. Here $r_{\rm gal}$ is the 3D half-mass radius of the galaxy, defined as the distance from the galaxy center of mass that encloses $50 \%$ of all star particles.

At $T=0$, by construction, the galaxy is at rest with respect to the soliton, and its center of mass coincides with the soliton center. Since the soliton is subject to both gravity and gradients in quantum pressure from the surrounding halo envelope, while the galaxy only feels the former, it is set in motion with respect to the soliton. As is evident from the bottom left-hand panel, the magnitude of this relative motion, characterized by $r_{\rm gs}$, is of the order of the soliton core radius, $r_\rmc$, indicated by the black, dashed, horizontal line. Since the galaxy is well embedded within the soliton at $T=0$ ($r_{\rm gal}=1.3 a_{\rm gal} \sim 0.4\ \kpc \sim 0.5 r_\rmc$), we see that $r_{\rm gs}/r_{\rm gal}$ is of order unity at early times (bottom right-hand panel), resulting in strong tidal perturbations throughout the galaxy. These tidal perturbations, combined with the order-unity density oscillations of the soliton, inject energy into the galaxy. This dynamical heating causes the velocity dispersion of the stars to increase (central right-hand panel). In turn, this increase in kinetic energy implies that the stars can venture out to larger radii than before, causing the galaxy to puff up, which manifests as an increase in its size and a decrease in the central density (top right-hand panel). 

The initial heating is very efficient, highlighted by the rapid increase in galaxy size and velocity dispersion at early times. However, as the galaxy continues to expand, the dynamical heating rate decreases. This reduced heating at later times has two causes. First, since the stars, on average, move further out from the soliton, the tidal heating induced by it reduces. Second, as the (central) velocity dispersion of the stars increases, they cross the soliton in a shorter time, making the potential fluctuations due to the soliton more adiabatic. Note that besides being perturbed by the soliton, the galaxy is also subject to additional perturbations from the density fluctuations (or quasiparticles) in the halo envelope. However, as the soliton is more massive than the quasiparticles (by at least an order of magnitude), it remains the most dominant source of perturbation at all times.

\begin{figure}
    \centering
    \includegraphics[width=0.45\textwidth]{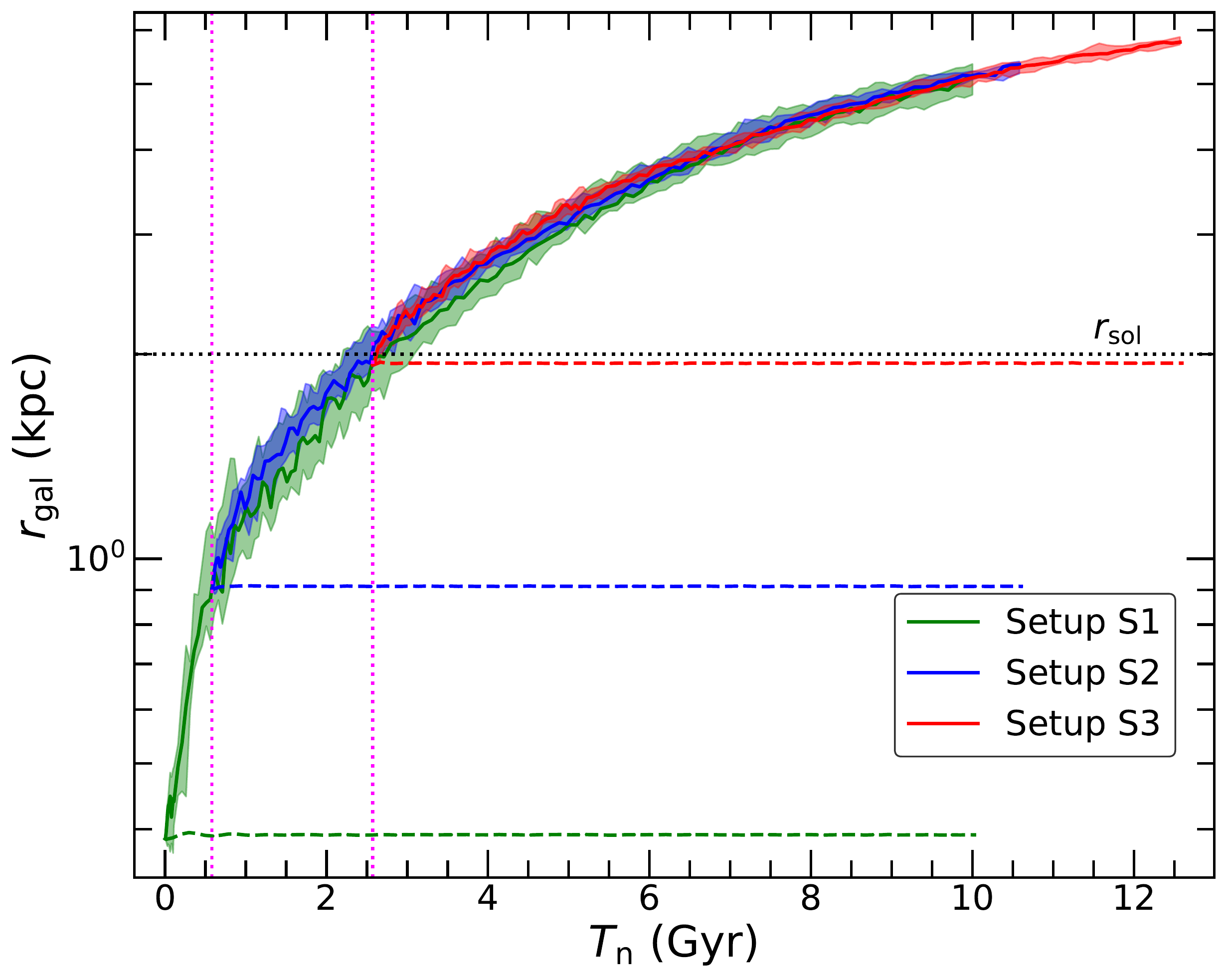}
    \caption{Evolution of the 3D half-mass radius, $r_{\rm gal}$, of the galaxies in Setups S1 (green), S2 (blue), and S3 (red). For each setup, the solid line and the associated envelope indicate, respectively, the median and the $16^{\rm th}-84^{\rm th}$ percentile variation over the 10 realizations in that setup. For comparison, the close-to-horizontal, dashed lines are the results obtained when the initial particle distributions in the three setups are evolved in the static, time-and-azimuthally averaged potential of the FDM halo. As is evident, without fluctuations, our simulated galaxies are perfectly stable. With fluctuations, though, $r_{\rm gal}$ increases steadily with time. Note that $r_{\rm gal}$ is plotted as a function of $T_\rmn = T + T_{\rm shift}$ (see text for details), which highlights that the outward diffusion of the stars due to the dynamical heating in the FDM halo is largely independent of the galaxy initial conditions. $T_{\rm shift}=0$ for Setup S1 and $0.6 \Gyr$ and $2.6 \Gyr$ for Setups S2 and S3, respectively, which are indicated by the vertical, dotted, magenta lines. The black, dotted, horizontal line marks the soliton radius, $r_{\rm sol}$.}
    \label{fig:halfmass}
\end{figure}

We conclude with a couple of interesting remarks. First, while the galaxy retains its overall spherical shape over time, its projected isodensity contours cease to be perfectly circular and concentric. This is discussed in detail in Section~\ref{sec:ellipse}. Second, while the galaxy starts out with an isotropic velocity distribution at $T=0$, over time, it develops a strong, radial anisotropy, especially in its outskirts. This is evident from the fact that at later times, $\sigma_{{\rm gal},r}(r) > \sigma_{{\rm gal}, \theta}(r) \sim \sigma_{{\rm gal}, \phi}(r)$ (central right-hand panel). The evolution in the kinematic structures of our simulated dwarf galaxies is discussed in detail in Section~\ref{sec:dispersion}. 

\begin{figure*}
    \centering
    \includegraphics[width=0.95\textwidth]{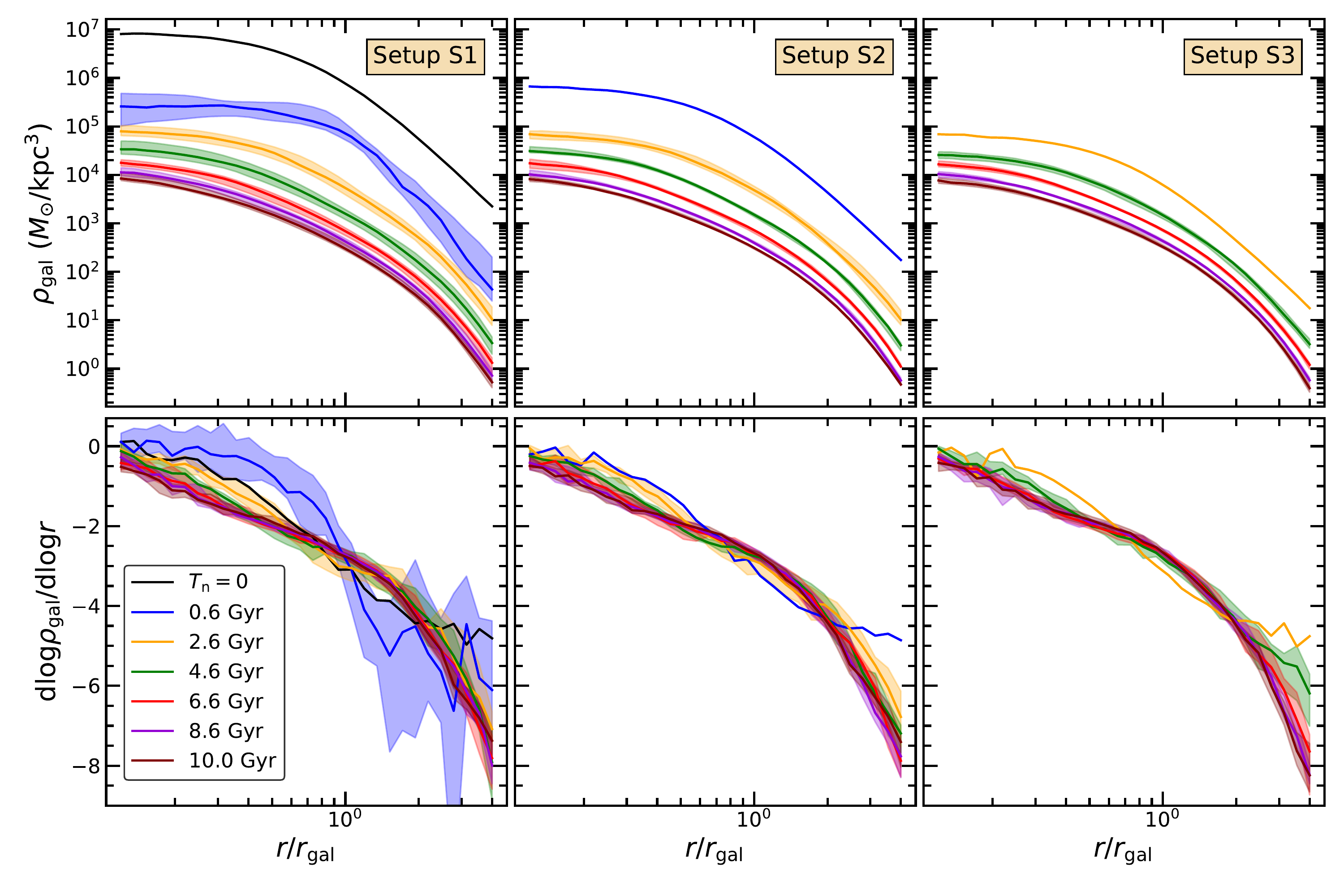}
    \caption{Evolution of the stellar density profiles. The top panels plot the azimuthally-averaged 3D density profiles of the galaxies. Different panels correspond to different setups, and different colors correspond to results at different epochs, $T_\rmn$, as indicated. Results are plotted as function of $r/r_{\rm gal}$, where $r_{\rm gal}$ is the half-mass radius {\it at that epoch}. The lower panels plot the corresponding logarithmic density gradients, $\rmd\log\rho_{\rm gal}/\rmd\log r$. As in Figure~\ref{fig:halfmass}, the solid lines and the associated envelopes indicate, respectively, the medians and the $16^{\rm th}-84^{\rm th}$ percentile variations over the 10 realizations of a given setup. Note how the density at a particular $r/r_{\rm gal}$ decreases with time, and the evolved profiles deviate significantly from the initial Plummer profiles.}
    \label{fig:density}
\end{figure*}

\subsection{Statistical Evolution of Key Properties}

We now describe the statistical evolution of some key properties of the dwarf galaxies in the different simulation setups. In order, we discuss the temporal evolution in the 3D half-mass radii, 3D density profiles, kinematic structures, and morphologies of our dwarf galaxies, focusing on the median and $1 \sigma$ variation over the different realizations of a particular setup.

\subsubsection{Half-Mass Radius}
\label{sec:half_mass}

Figure~\ref{fig:halfmass} shows the evolution in $r_{\rm gal}$ of the dwarf galaxies in Setups S1 (green), S2 (blue), and S3 (red). For each setup, the solid line and the associated envelope indicate, respectively, the median and the $16^{\rm th}-84^{\rm th}$ percentile variation over the 10 realizations in that setup. Note that the results are plotted as a function of $T_\rmn=T+T_{\rm shift}$. Here, $T_{\rm shift}=0$ for Setup S1, and for Setups S2 and S3, it is the time taken for the median $r_{\rm gal}$ from Setup S1 to attain a value equal to the initial $r_{\rm gal}$ in these setups ($\sim 0.9 \kpc$ and $\sim 2 \kpc$, respectively). The vertical, dotted, magenta lines indicate the values of $T_{\rm shift}$, which are $0.6\Gyr$ and $2.6\Gyr$, for Setups S2 and S3, respectively. For comparison, the initial particle distributions in all three setups are also evolved for $10\Gyr$ in the fixed, time-and-azimuthally averaged halo potential, $\Phi_\rmh$, and the corresponding results are shown with the dashed curves. 

In the latter case, due to the absence of any potential fluctuations, the half-mass radii of the galaxies, which are set up to be in equilibrium in $\Phi_\rmh$, remain constant in time. However, in a live FDM halo, the potential fluctuations arising from the random walk of the soliton, its oscillations, and the density fluctuations in the halo envelope continuously perturb and dynamically heat the dwarf galaxies, causing $r_{\rm gal}$ to increase over time. The black, dotted, horizontal line indicates the soliton radius, $r_{\rm sol}$, and the galaxies in Setups S1 and S2, which are initially smaller than the soliton, increase their median $r_{\rm gal}$ to $r_{\rm sol}$ in $\sim 2.6$ and $\sim 2 \Gyr$, respectively.

Note that the rate at which $r_{\rm gal}$ increases is smaller for larger $r_{\rm gal}$, which is a consequence of the reduced heating efficiency in larger systems, as discussed in Section~\ref{sec:example}. Also, note that the evolution of $r_{\rm gal}$ in all three setups roughly follows the same trend, independent of the different initial conditions. This trend is well described with a power law, $r_{\rm gal} \propto T^{\gamma}_\rmn$, but the index $\gamma$ transitions from $\simeq 0.5$ when $r_{\rm gal} \lesssim r_{\rm sol}$ (characteristic of Brownian motion) to $\simeq 0.7$ when $r_{\rm gal} \gtrsim r_{\rm sol}$. Using these relations, we can also estimate the time taken for a galaxy to double its size, which is shorter for smaller $r_{\rm gal}$. For example, the doubling time increases from $\sim 0.4\Gyr$ when $r_{\rm gal} = 0.4\kpc$ to $\sim 17 \Gyr$ when $r_{\rm gal} = 5 \kpc$.

\begin{figure*}
    \centering
    \includegraphics[width=0.95\textwidth]{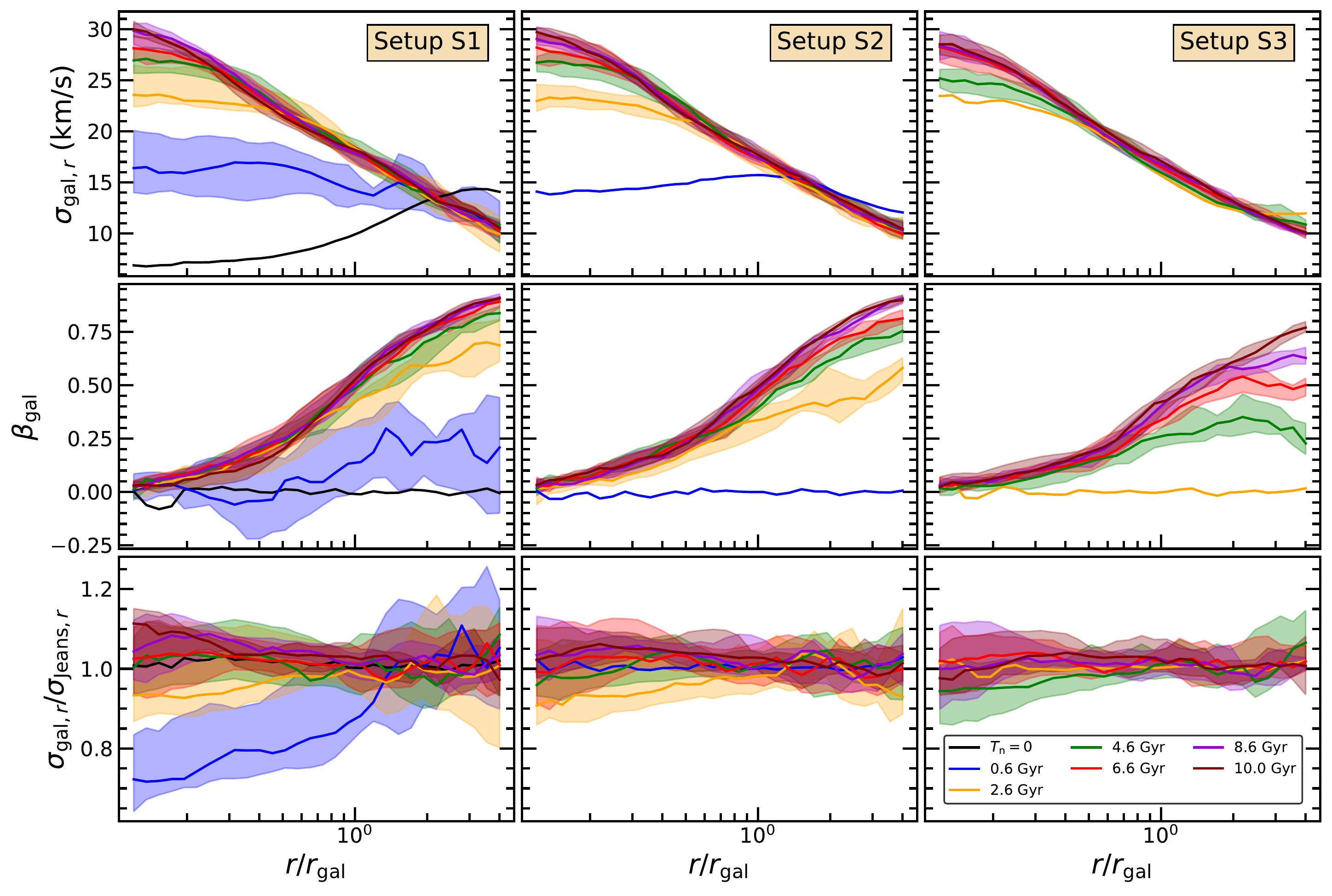}
    \caption{Same as Figure~\ref{fig:density} except that here we plot the profiles of the radial velocity dispersion, $\sigma_{{\rm gal},r}$ (upper panels), the velocity anisotropy parameter, $\beta$ (middle panels), and the ratio of $\sigma_{{\rm gal},r}$ and the radial velocity dispersion profile that satisfies the radial Jeans equations, $\sigma_{{\rm Jeans},r}$ (see text for details). Note how, over time, the radial velocity dispersion increases, the galaxies become highly radially anisotropic in their outskirts, and $\sigma_{{\rm gal},r}/\sigma_{{\rm Jeans},r}$ is close to unity. The latter indicates that despite the ongoing dynamical heating, the galaxies are in quasi-equilibrium.}
    \label{fig:dispersion}
\end{figure*}

\subsubsection{Density Profile}
\label{sec:density}

Figure~\ref{fig:density} depicts the evolution in the density profiles of our simulated galaxies, with the left-hand, middle, and right-hand panels corresponding to Setups S1, S2, and S3, respectively. The top row plots $\rho_{\rm gal}$ as a function of $r$ (distance from the galaxy center of mass), normalized by $r_{\rm gal}$, for specific values of $T_\rmn$, as indicated. The bottom row shows the corresponding logarithmic density gradients, $\rmd {\rm log} \rho_{\rm gal}/ \rmd {\rm log}r$. Recall that at a particular $T_\rmn$, the galaxies in the different setups have roughly the same size (see Section~\ref{sec:half_mass}).

Due to the dynamical heating, the galaxies puff up, causing the stellar densities at a given $r/r_{\rm gal}$ to decrease monotonically with time. Note that the evolution in $\rho_{\rm gal}$ slows down as the galaxies become larger (i.e., at later $T_n$), in agreement with the temporal trend of $r_{\rm gal}$ (cf.~Figure~\ref{fig:halfmass}). For example, in Setup S1, the peak density drops by a factor of $\sim 100$ in the first $2.6 \Gyr$ (black to orange) but only by a factor of $\sim 10$ during the next $7.4 \Gyr$ (orange to brown). The bottom row highlights that the evolved stellar density profiles are significantly different in shape from the initial Plummer profiles. In particular, after a few Gyr of dynamical heating, the logarithmic density slope as a function of $r/r_{\rm gal}$ settles into a broken power-law, decreasing monotonically outward, from close to zero in the center to $\sim -3$ at $r_{\rm gal}$ to $< -7$ for $r > 3.5 r_{\rm gal}$. Hence, the outskirts of the heated dwarfs are significantly steeper than the initial Plummer profiles, which have  $\rmd {\rm log} \rho_{\rm gal}/ \rmd {\rm log}r \to -5$ for $r/r_{\rm gal} \to \infty$.

\subsubsection{Kinematic Structure}
\label{sec:dispersion}

Figure~\ref{fig:dispersion} shows the evolution of various quantities related to the kinematic structures of our simulated galaxies, with the left-hand, middle, and right-hand panels corresponding to Setups S1, S2, and S3, respectively. From top to bottom, the three different rows plot the radial velocity dispersion, $\sigma_{{\rm gal},r}$, the velocity anisotropy parameter,
\begin{equation}
\beta_{\rm gal}(r) = 1-\frac{\sigma^{2}_{{\rm gal}, \theta}(r)+\sigma^{2}_{{\rm gal},\phi}(r)}{2 \sigma^{2}_{{\rm gal},r}(r)} \,, 
\end{equation}
and the ratio $\sigma_{{\rm gal},r}(r)/\sigma_{{\rm Jeans},r}(r)$, respectively, all as a function of $r/r_{\rm gal}$ and for specific values of $T_\rmn$, as indicated. Here, $\sigma_{{\rm Jeans},r}(r)$ is the radial velocity dispersion profile that solves the radial Jeans equation
\begin{equation}
\frac{1}{\rho_{\rm gal}} \frac{\rmd \rho_{\rm gal} \sigma^2_{{\rm Jeans},r}}{\rmd r} + 2 \beta_{\rm gal} \frac{\sigma^2_{{\rm Jeans},r}}{r} = -\frac{\rmd \Phi_\rmh}{\rmd r} \,,
\end{equation}
where $\Phi_\rmh(r)$ is the time-and-azimuthally averaged potential of the FDM halo, and $\rho_{\rm gal}(r)$ and $\beta_{\rm gal}(r)$ are the actual density and velocity anisotropy profiles (azimuthally averaged) at time $T_\rmn$ taken from the simulations. Note that in this calculation, we ignore the self-gravity of the stars (which is anyways negligible) and assume that the galaxy center of mass coincides with the soliton center. 

From the top panels, it is clear that the dynamical heating induced by the potential perturbations in the FDM halo causes $\sigma_{{\rm gal},r}$ to increase with time. The heating is more efficient at small $T_\rmn$, when the galaxies are smaller in size and kinematically colder. At large $T_\rmn$, the radial velocity dispersion profiles, expressed as a function of $r/r_{\rm gal}$, seem to become time-invariant. Note, though, that $r_{\rm gal}$ itself continues to increase, and thus that dynamical heating is ongoing, i.e., at a fixed $r$, the velocity dispersion continues to increase with time.

The middle panels show that, starting from an isotropic velocity distribution ($\beta_{\rm gal}(r)=0$), our simulated galaxies become radially anisotropic ($\beta_{\rm gal}(r)>0$) over time, especially in their outskirts, reminiscent of Osipkov-Merrit models \citep[][]{osipkov79, merritt85}. In fact, at late times, $\beta_{\rm gal}$ increases from zero near the center to $> 0.75$ for $r > 2r_{\rm gal}$. Hence, the outskirts of dwarf galaxies in FDM halos are predicted to have quite extreme radial anisotropies. This can be understood from the fact that the central soliton is the most dominant heating source. As a result, stars receive their largest velocity impulses during their pericentric passages, causing their orbits to become more eccentric over time. Consequently, stars at large galactocentric radii are likely to be near their apocenters, which explains why the outskirts are more radially anisotropic.

\begin{figure}
    \centering
    \includegraphics[width=0.45\textwidth]{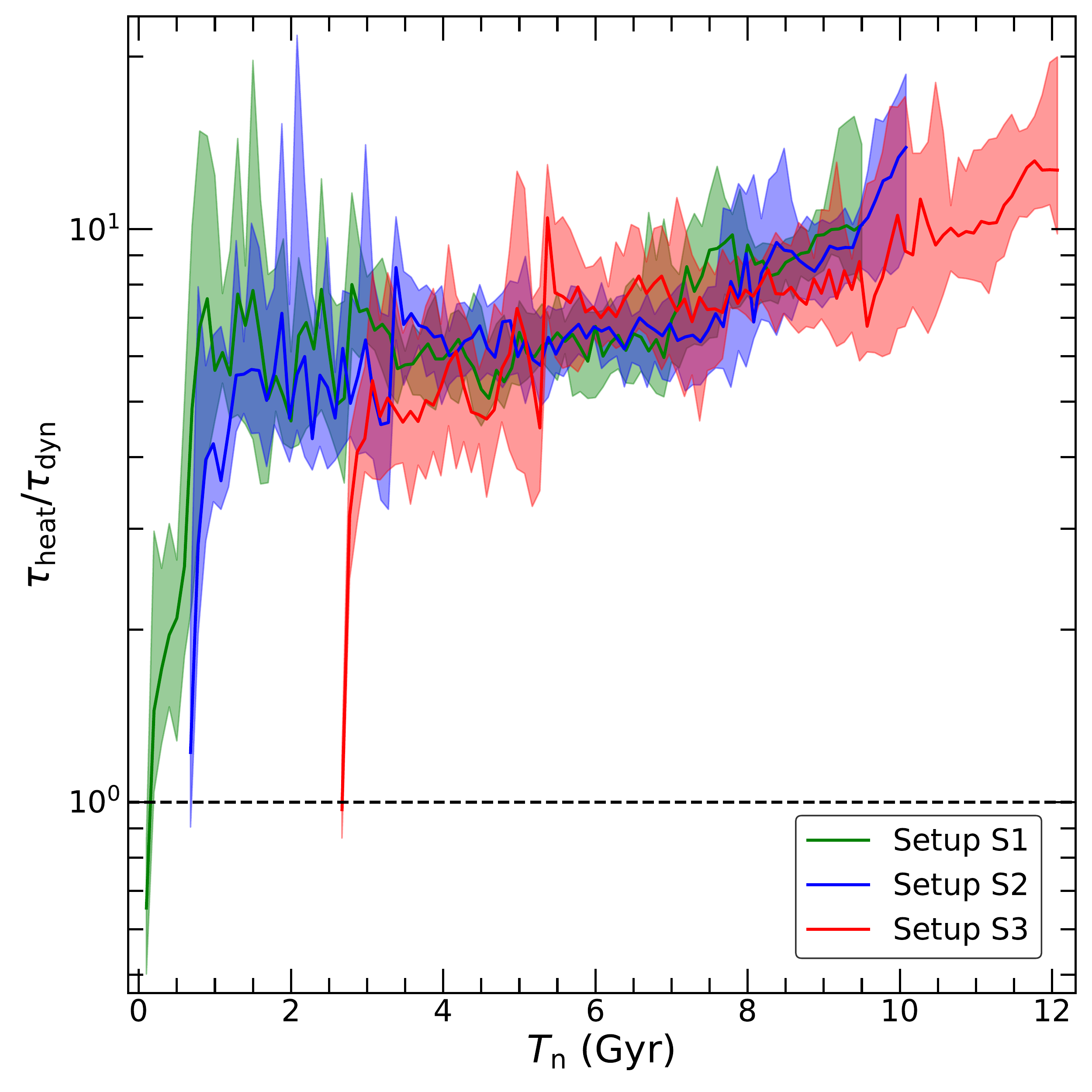}
    \caption{The ratio of the heating time scale, $\tau_{\rm heat}$ (Equation~\ref{tauheat}) to the dynamical time scale, $\tau_{\rm dyn}$ (Equation~\ref{taudyn}) as a function of $T_\rmn$. Results are shown for Setups S1 (green), S2 (blue), and S3 (red), with solid lines and associated envelopes indicating the median and the $16^{\rm th}-84^{\rm th}$ percentile variation over the 10 different realizations in a particular setup. Note that throughout most of the evolution, this ratio is well in excess of unity, indicating that despite the ongoing dynamical heating, the galaxies are in quasi-equilibrium.}
    \label{fig:timescales}
\end{figure}

Finally, the bottom panels reveal that except for small $T_{\rmn}$ ($T_{\rmn}\lta 1 \Gyr$),  $\sigma_{{\rm gal},r}$ is consistent (within $\pm 10 \%$) with $\sigma_{{\rm Jeans},r}$ at all $r$, indicating that the galaxies are not very far off from equilibrium. This implies that the timescale for dynamical heating must be large compared to the dynamical timescale, such that the system has sufficient time to equilibrate to a slowly changing quasi-equilibrium state. Figure~\ref{fig:timescales} shows that this is indeed the case. It plots the time evolution of the ratio of the dynamical heating time,
\begin{equation}
\tau_{\rm heat}=\frac{r_{\rm gal}}{\rmd r_{\rm gal}/ \rmd t} \,, 
\label{tauheat}
\end{equation}
to an estimate of the dynamical time,
\begin{equation}
\tau_{\rm dyn} = \frac{2 \pi r_{\rm gal}}{ \sigma_{{\rm gal},r}(r_{\rm gal})} \,.
\label{taudyn}
\end{equation}
As is evident, throughout most of the evolution, this ratio is well in excess of unity. Hence, despite the ongoing dynamical heating, our simulated dwarf galaxies are to a good approximation in equilibrium and adequately described by (anisotropic) Jeans models.

\subsubsection{Morphology}
\label{sec:ellipse}

Having discussed how over time, the dynamical heating inside FDM halos causes dwarf galaxies to become extended, kinematically hot, radially anisotropic systems in quasi-equilibrium, we now focus on the evolution of their morphologies, as characterized by their projected isodensity contours. Here, we also compare the evolution of the initially spherical galaxies to the ones that are initially ellipsoidal (see Section~\ref{galaxy_ics}).

\begin{figure*}
    \centering
    \includegraphics[width=0.95\textwidth]{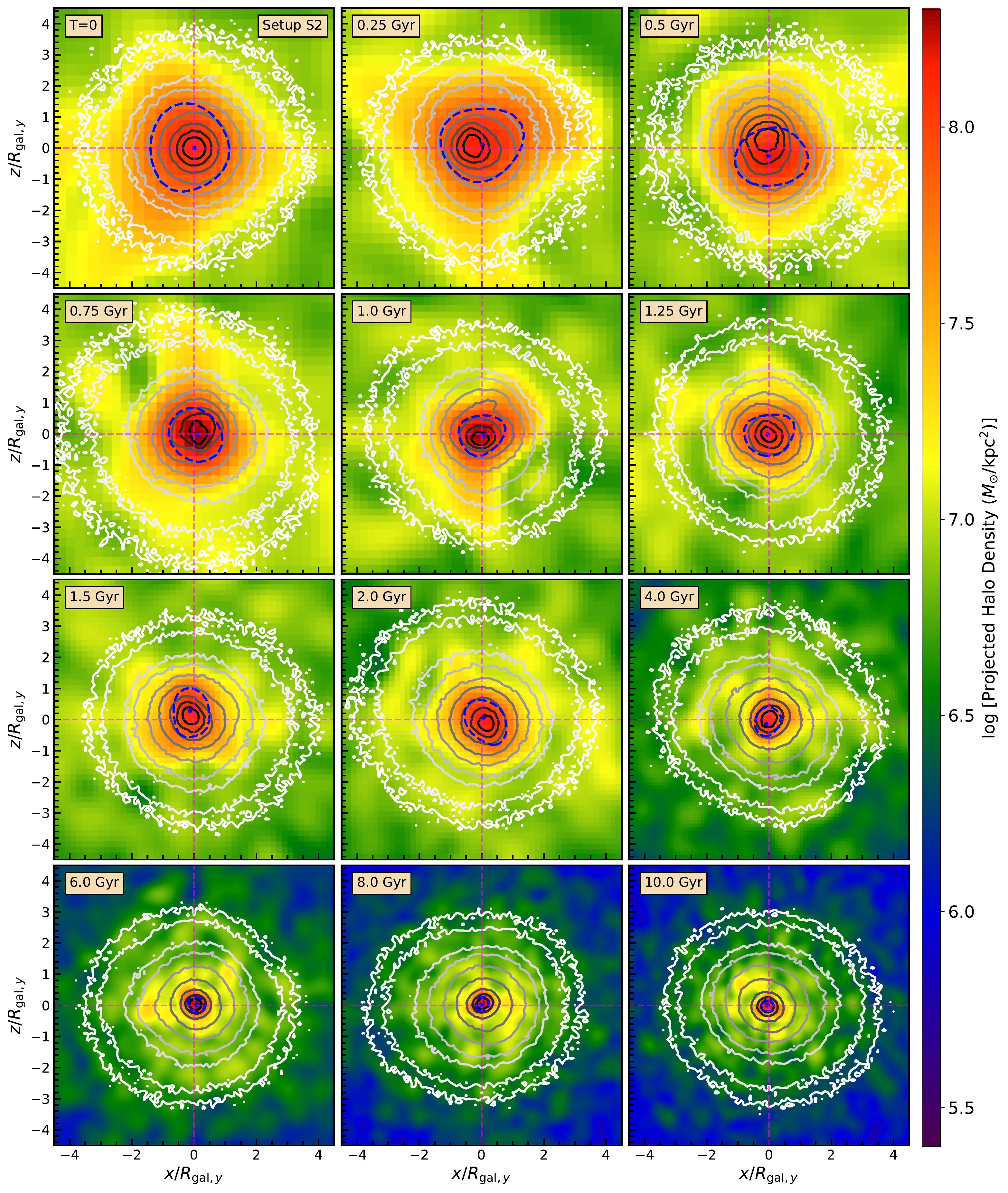}
    \caption{Morphological evolution of the dwarf galaxy in one of the realizations of Setup S2. In each panel, corresponding to different epochs, as indicated, the color map shows the dark matter density projected onto the $x$-$z$ plane (see the color bar for scale), while the solid curves indicate the isodensity contours of the projected stellar surface density. The $x$ and $z$ positions are in the center of mass frame of the stellar body (indicated by the dashed, magenta cross-hairs) and normalized by the projected stellar half mass radius in the same frame, $R_{{\rm gal},y}$, to do away with the galaxy's overall expansion. As the galaxy expands with time, more and more of the outer envelope of the FDM halo becomes visible. The blue dot marks the pixel with the highest FDM density, indicative of the soliton center in projection, and the blue, dashed curve marks the isodensity contour having a value equal to half of the maximum projected halo density, roughly outlining the extent of the solitonic core. As the galaxy is continuously perturbed by the FDM potential fluctuations, its isophotes deviate from perfect sphericity and cease to be concentric (by construction, they are spherical and concentric at $t=0$).}
    \label{fig:image_halo_spherical}
\end{figure*}

\begin{figure*}
    \centering
    \includegraphics[width=0.95\textwidth]{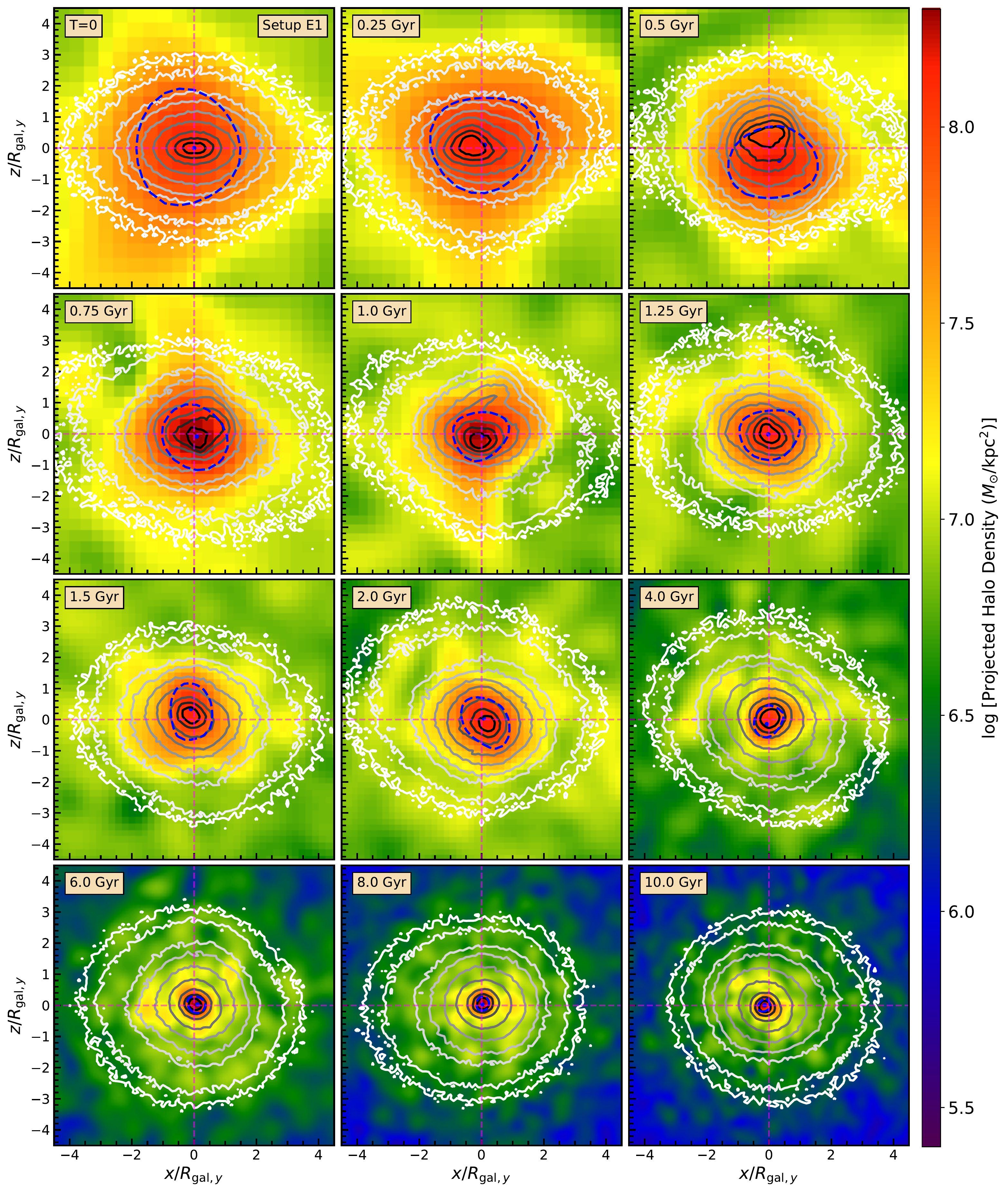}
    \caption{Same as Figure~\ref{fig:image_halo_spherical} except that here we show the results for one of the realizations of Setup E1, for which the initial galaxy is significantly flattened along the $z$-axis (cf. Figure~\ref{fig:initial_ellipticity}). Initially elliptical galaxies become roughly spherically symmetric over time from the inside out. However, as the galaxy is continuously perturbed by the FDM potential fluctuations, the isophotes always deviate from perfect sphericity and cease to be concentric.}
    \label{fig:image_halo_elliptical}
\end{figure*}

For a randomly selected realization in Setup S2, Figure~\ref{fig:image_halo_spherical} depicts the evolution in the 2D shape of its resident dwarf galaxy, projected along the $y$-direction. Different panels correspond to different times, as indicated. In each image, centered on the instantaneous center of mass of the galaxy (magenta cross-hairs), the color map shows the projected density of the FDM halo, while the solid curves highlight the projected isodensity contours of the galaxy. The galaxy contours vary from $0.8 \Sigma_{\rm gal,m}$ (black) to $0.005 \Sigma_{\rm gal,m}$ (white), where $\Sigma_{\rm gal,m}$ is the maximum projected stellar surface density at a given time. Note that the spatial range in all panels runs from $-4 R_{{\rm gal},y}$ to $4 R_{{\rm gal},y}$, where $R_{{\rm gal},y}$ is the distance from the galaxy center of mass that encloses $50 \%$ of all star particles in the $y$-projection. Hence, as the galaxy expands with time, more and more of the outer halo envelope beyond the soliton becomes visible, and the soliton core, whose extent is indicated by the dashed, blue contour, appears to shrink. Figure~\ref {fig:image_halo_elliptical} shows the same as Figure~\ref{fig:image_halo_spherical} but for a randomly selected realization belonging to Setup E1, in which the galaxy starts out with an initial flattening along the $z$-axis (see Figure~\ref{fig:initial_ellipticity}).

During the first $1-2 \Gyr$, tidal distortions due to the wobbling soliton cause the central regions of the galaxies to undergo strong asymmetric distortions, and the isodensity contours clearly are no longer concentric. Over time, though, as the galaxies puff up, each isodensity contour that encloses a fixed fraction of the stars moves out and becomes more roundish (on average). This owes to the fact that integrated over a sufficiently long time, the net heating is isotropic.

\begin{figure*}
    \centering
    \includegraphics[width=0.95\textwidth]{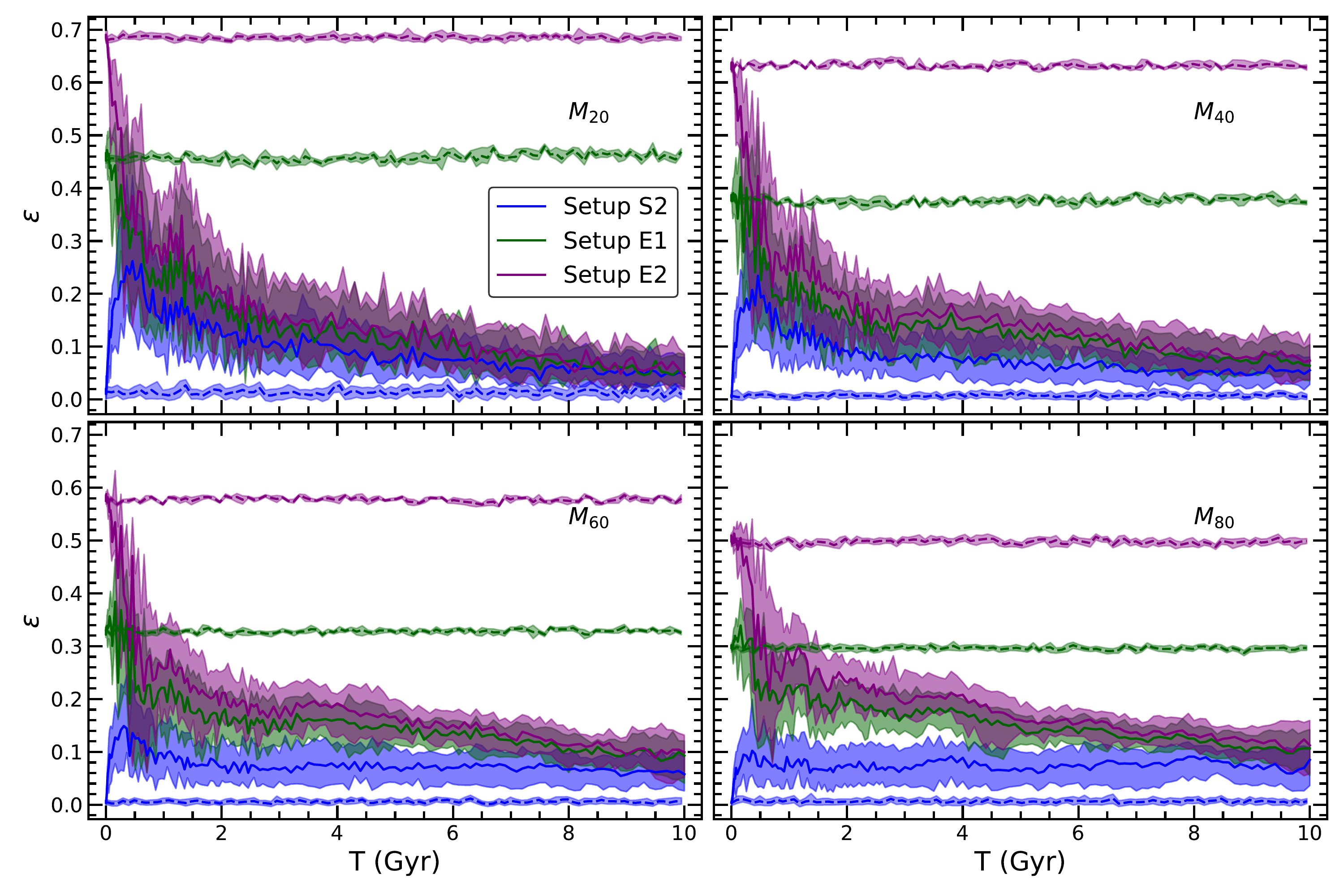}
    \caption{Evolution of the ellipticity, $\varepsilon$, of the isodensity contours that enclose $20 \%$ (top left), $40 \%$ (top right), $60 \%$ (bottom left), and $80 \%$ (bottom right) of the total galaxy mass in projection. Results are shown for Setups S2 (blue), E1 (dark green), and E2 (magenta). The solid curves and the associated envelopes, respectively, indicate the medians and $16^{\rm th}-84^{\rm th}$ variations in $\varepsilon$ over the 10 different realizations of a particular setup. In addition, the dashed curves and the associated envelopes indicate the results obtained when the initial particle distributions are evolved in the static, time-and-azimuthally averaged halo potential. As is evident, without fluctuations, the initial ellipticity is perfectly preserved, whereas the FDM fluctuations drive the system towards an almost spherical symmetry, independent of the initial flattening. Hence, if dark matter is indeed fuzzy, with $m_\rmb \sim 10^{-22} \eV$, dwarf galaxies with old stellar populations are predicted to be close to spherical.}
    \label{fig:ellipticity}
\end{figure*}

\begin{figure*}
    \centering
    \includegraphics[width=0.95\textwidth]{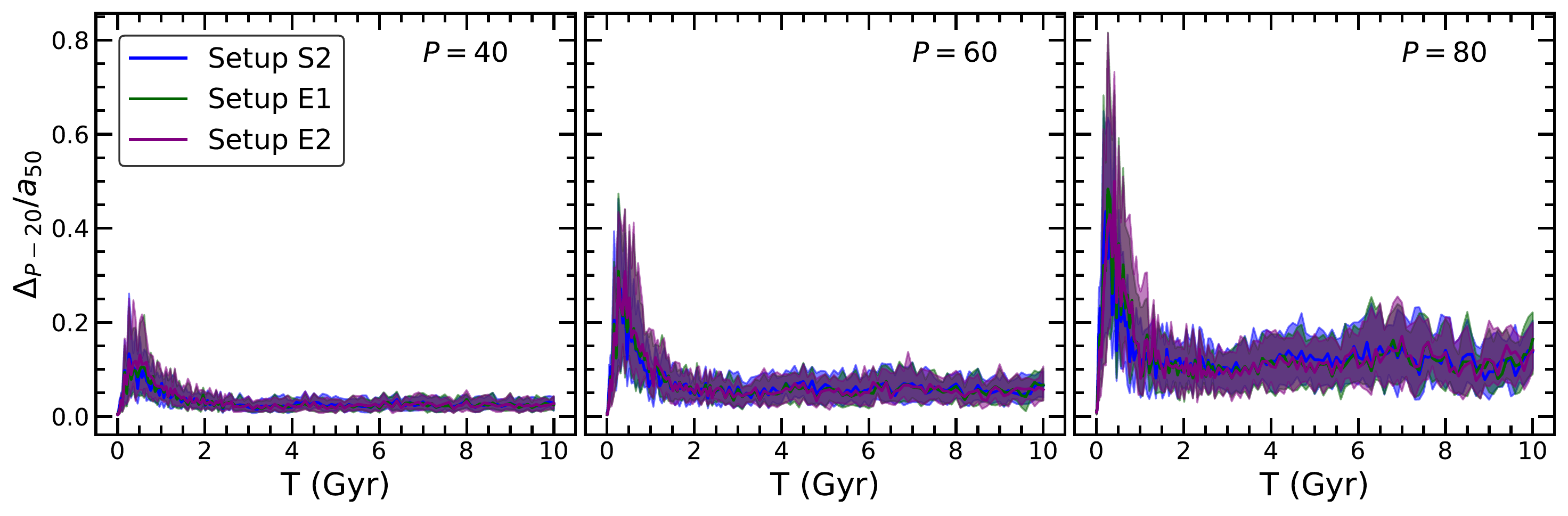}
    \caption{Evolution of the normalized offset parameter, $\Delta_{P-20}/a_{50}$, between the centers of the ellipses that best fit the isodensity contours that enclose $P\%$ and $20\%$ of the stellar mass in projection. Here, $a_{50}$ is the semi-major axis of the isodensity contour corresponding to $P=50$. Results are shown for $P=40$ (left-hand panel), $P=60$ (middle panel), and $P=80$ (right-hand panel) and for Setups S2 (blue), E1 (dark green), and E2 (purple). The solid curves and the associated envelopes, respectively, indicate the medians and $16^{\rm th}-84^{\rm th}$ percentile variations over the 10 different realizations of a particular setup. Note how tidal distortions caused by the soliton initially create large asymmetries (resulting in large offsets). At later times, though, $\Delta_{P-20}/a_{50}$ roughly asymptotes to a constant value, which increases with $P$. Since $a_{50}$ continues to increase, the offsets $\Delta_{P-20}$ continue to increase as well in an absolute sense. Hence, for $m_{\rmb} \sim 10^{-22} \eV$, dwarf galaxies in FDM halos are predicted to have isophotes that are significantly non-concentric.}
    \label{fig:offset_evolve}
\end{figure*}

In order to quantify these trends, we fit the isodensity contours of the $y$-projections of all our simulated galaxies with ellipses using the {\it photutils} package \citep{bradley20} and track the evolution in the ellipticities, $\varepsilon$, and the centers of the isodensity contours over time. Figure~\ref{fig:ellipticity} shows the evolution in $\varepsilon$ of the isodensity ellipses that enclose $20 \%$ ($M_{20}$, top left), $40 \%$ ($M_{40}$, top right), $60 \%$ ($M_{60}$, bottom left), and $80 \%$ ($M_{80}$, bottom right) of the total galaxy mass in projection for Setups S2 (blue), E1 (dark green), and E2 (magenta). At early times, the tidal heating due to the soliton introduces some ellipticity within the initially spherical galaxies of Setup S2, causing $\varepsilon$ to grow with time for $\sim 0.5\Gyr$, reaching larger values for isodensity contours that enclose smaller mass fractions. Thereafter, the ellipticities decline to a value of $\sim 0.05$ with very little dependence on the enclosed mass. For Setups E1 and E2, the initial ellipticity declines rapidly, and the galaxies becomes more spherical over time from the inside out. This radial trend is simply a consequence of the fact that the dynamical heating, which isotropizes the galaxies, is stronger at smaller radii. After about $8 \Gyr$ of evolution, the galaxies in Setups E1 and E2 attain the same level of flattening as that in Setup S2; i.e., the dynamical heating has erased the initial conditions. Hence, if dark matter is indeed fuzzy and $m_\rmb \sim 10^{-22} \eV$, then dwarf galaxies with old stellar populations should all be close to spherical.

As mentioned above, the wobbling of the soliton also causes strong asymmetries in the isodensity contours, especially those whose semi-major axes are small or comparable to the soliton radius. As a consequence, the isodensity contours are not all concentric. Figure~\ref{fig:offset_evolve} shows the evolution of $\Delta_{P-20}$, defined as the offset between the centers of $M_{P}$ and $M_{20}$, normalized by $a_{50}$ (the semi-major axis of $M_{50}$). Results are shown for Setups S2 (blue), E1 (dark green), and E2 (purple), and the left-hand, middle, and right-hand panels correspond to $P=40$, $60$, and $80$, respectively. For all setups, the normalized offset parameter, $\Delta_{P-20}/a_{50}$, initially increases with time for about $0.5 \Gyr$, after which it decreases again, asymptoting to a constant value after roughly $2 \Gyr$. The asymptotic value is independent of the initial ellipticity of the galaxy, but increases slightly with $P$, from $0.03 \pm 0.01$ for $P=40$ to $0.06 \pm 0.02$ for $P=60$ to $0.12 \pm 0.06$ for $P=80$. Hence, if dark matter is indeed fuzzy and $m_\rmb \sim 10^{-22} \eV$, then the isophotes of old dwarf galaxies that enclose $20$ and $80 \%$ of the total (projected) stellar light should (on average) be offset from each other by about $10 \%$ of the half-light radius. 

\section{Summary and Conclusion} \label{sec:concl}

We studied the evolution of dwarf galaxies in an isolated FDM halo, assuming that dark matter is made up of ultralight bosons with mass, $m_\rmb = 8 \times 10^{-23}\eV$. Using the code {\tt GAMER-2}, which solves the SP equation, we evolved galaxies of mass $M_{\rm gal}=10^{6} M_{\odot}$, consisting of $10^6$ particles, in a live FDM halo with a virial mass of $6.6 \times 10^{9} M_{\odot}$. The galaxies differ in their initial size and/or flattening and are initialized to be in equilibrium in the time-and-azimuthally averaged halo potential. However, in a live FDM halo, they are subject to persistent dynamical heating, which has a drastic impact on their structural and kinematic properties. In particular, we have shown that

\begin{itemize}
    \item The heating causes the velocity dispersions of our simulated dwarf galaxies to increase with time at all radii but more so closer to the center. As a consequence, their half-mass radii continuously increase with time as well, at a rate that roughly follows $\rmd\log r_{\rm gal}/\rmd\log t = 0.7$ ($0.5$) when $r_{\rm gal}$ is larger (smaller) than the soliton radius, $r_{\rm sol}$. 
    
    \item As the galaxies expand, their velocity distributions become strongly radially anisotropic, especially in the outskirts. This is a consequence of the heating being most pronounced in the central regions of the halo, where the wobbling, oscillating soliton imparts the stars with large velocity impulses.
    
    \item Since the long-term impact of the dynamical heating is isotropic, the galaxies become roughly spherically symmetric over time from the inside out, independent of their initial shape. Note, though, that as long as the heating continues, the shapes will always deviate somewhat from perfect sphericity. 
    
    \item The isophotes (or isodensity contours) of the galaxies deviate significantly from being concentric. In particular, we find that, at late times, the offset between the centers of the isophotes that enclose $20$ and $80 \%$ of the total (projected) stellar light is roughly $10 \%$ of the galaxy's half-light radius (on average). In addition, due to strong tidal distortions caused by the soliton, isophotes with semi-major axes that are small or comparable to $r_{\rm sol}$ typically show strong asymmetries (i.e., are poorly fit by simple ellipses).
    
    \item As long as the half-mass radii of the galaxies are larger than or comparable to the core radius of the soliton, the heating time scales are significantly longer than the dynamical time scales, such that the galaxies are to a good approximation in equilibrium. In particular, they can be adequately described using anisotropic Jeans models.
\end{itemize}

Clearly, then, the dynamical heating caused by the gravitational potential perturbations in FDM halos has a profound impact on dwarf galaxies, and it is tempting to use the results outlined above to proclaim that FDM would be inconsistent with the presence of significantly flattened dwarf galaxies with (close to) concentric isophotes or to argue that FDM models predict that (dwarf) galaxies should obey an age-size relation. However, we emphasize that the results presented here have only focused on a single boson mass ($m_\rmb = 8 \times 10^{-23}\eV$) and a single FDM halo (with virial mass of $\Mvir \simeq 6.6 \times 10^{9} M_{\odot}$) in isolation. In addition, the galaxies have been modeled as single-age stellar populations. Making more meaningful predictions for the statistical properties of dwarf galaxies for a given FDM model will require exploring a wider range in halo masses, accounting for merger histories, and for extended star formation histories of their galaxies. The results presented here only serve to illustrate that (i) dynamical heating inside dark matter halos is a key prediction of FDM models, and (ii) dynamical heating leaves several characteristic signatures in the population of dwarf galaxies. 

Self-consistent FDM simulations, such as the ones presented here, are expensive (particularly for larger boson masses). Therefore, in Dutta Chowdhury et al. (in preparation), we construct a semi-analytical model (calibrated against a few simulations) to estimate the dynamical heating in FDM for any halo and boson mass. This model can then be used to make predictions for the key dwarf galaxy observables discussed in this paper (size, velocity dispersion and anisotropy, shape and isophotal offset) as a function of both halo and boson mass. Such predictions can be compared to data of isolated dwarf galaxies to be obtained with upcoming surveys such as the \textit{Legacy Survey of Space and Time} \citep[LSST,][]{brough20, kaviraj20} and the \textit{Dragonfly Wide Field Survey} \citep[][]{danieli20} in an attempt to constrain the FDM boson mass.

\section*{Acknowledgments}
The authors thank the anonymous referee for insightful feedback that helped in improving the manuscript. DDC is grateful to Uddipan Banik, Neal Dalal, Avishai Dekel, Nir Mandelker, and Kaustav Mitra for valuable discussions and thanks the Yale Center for Research Computing for guidance and use of the research computing infrastructure, specifically the Grace cluster. FvdB is supported by the National Aeronautics and Space Administration through Grant No. 19-ATP19-0059 issued as part of the Astrophysics Theory Program and received additional support from the Klaus Tschira foundation. V.H.R. acknowledges support by YCAA Prize postdoctoral fellowship. H. S. acknowledges funding support from the Jade Mountain Young Scholar Award No. NTU-110V0201, sponsored by the Ministry of Education, Taiwan. This research is partially supported by the Ministry of Science and Technology (MOST) of Taiwan under Grants No. MOST 107-2119-M-002-036-MY3 and No. MOST 108-2112-M-002-023-MY3, and the NTU Core Consortium project under Grants No. NTU-CC-108L893401 and No. NTU-CC-108L893402.

\bibliography{ms}{}
\bibliographystyle{aasjournal}
\end{document}